\documentclass[prd,aps,reprint,onecolumn,superscriptaddress,tightenlines,nofootinbib,eqsecnum,preprintnumbers,longbibliography,12pt]{revtex4-2}

\usepackage{epsfig,amsfonts,mathrsfs,amsmath,amssymb,graphicx,color,slashed}
\usepackage{amsmath,latexsym,amssymb,graphicx,slashed,hyperref,enumerate,url,etoolbox,cancel,gensymb,physics}
\hypersetup{colorlinks,citecolor= nicegreen,linkcolor= nicered}
\usepackage[usenames,dvipsnames]{xcolor}
\definecolor{nicered}{rgb}{0.7,0.1,0.1}
\definecolor{nicegreen}{rgb}{0.1,0.5,0.1}
\definecolor{niceblue}{rgb}{0.1,0.2,0.6}

\usepackage{bbm}
\usepackage{subfig}
\usepackage{physics}
\usepackage{dsfont}

\def\Carleton{Ottawa-Carleton Institute for Physics, Carleton University, Ottawa, ON K1S 5B6, Canada}
\def\UCSD{Department of Physics, University of California, San Diego, CA 92093, USA}

\begin{document}


\title{Exploring the vacuum structure of gravitationally induced neutrino masses
}
\def\Carleton{Ottawa-Carleton Institute for Physics, Carleton University, Ottawa, ON K1S 5B6, Canada}

\author{Carlos Henrique de Lima}
\email{carloshenriquedelima@cmail.carleton.ca}
\affiliation{\Carleton}
\author{Daniel Stolarski}
\email{stolar@physics.carleton.ca}
\affiliation{\Carleton}
\affiliation{\UCSD}

\date{\today}
\begin{abstract}
In this work, we explore the proposed mechanism in which the gravitational $\theta$ anomaly generates neutrino masses. We highlight that the leading renormalizable interactions of the neutrino condensate forbid the possibility of generating hierarchical masses consistent with observation. This conclusion still holds when Standard Model loop corrections are accounted for. We show that higher-dimensional operators can alleviate this problem. The higher-dimensional operators could be generated from the gravitational anomaly itself, but there is no clear way to know without a deeper understanding of the low-energy description of this mechanism. Because of that, we explore the possibility of new particles generating neutrino mass splittings. We show that both new particles that alter the scalar potential of the condensate or new particles in loops for the neutrino self-energy can solve this problem.
\end{abstract}


\maketitle

\section{Introduction}
The underlying mechanism that generates neutrino masses is still an open question in physics since the measurement of neutrino oscillations~\cite{Super-Kamiokande:1998kpq,oscEXP2}. Most scenarios proposed to generate neutrino masses include new physics at high energies~\cite{neutrinoREV}. Another direction which is promising is the condensation of neutrinos~\cite{cond1,cond2,cond3,cond4,cond5,Barenboim:2019fmj}. Usually, the condensation mechanism occurs at high energies and needs new fields. An alternative approach was proposed in~\cite{gravtheta} and reformulated in~\cite{Dvali:2017mpy}, in which low-energy physics generates neutrino masses.

This low energy generation of neutrino masses requires that, in pure gravity, we have a non-zero topological vacuum susceptibility at zero momentum. This condition is analogous to requiring the QCD $\theta$-term to be physical~\cite{anomaly1,anomaly2,anomaly3} such that the anomaly explicitly breaks the axial symmetry and gives mass to the $\eta'$ meson. Depending on whether there is a right-handed neutrino, the mechanism can generate a Dirac or Majorana mass for the neutrinos from the anomaly on the gravitational $\theta$-term. Phenomenological bounds give the scale of the condensate to be around the $\text{meV}$ range~\cite{gravtheta}. 

Naively, gravitational interactions would generate a universal mass for the three neutrinos. Of course, that is not what is observed in nature, with the mass squared splittings measured to be~\cite{Super-Kamiokande:2017yvm,PDG} 
\begin{align}\label{eq:splitting}
\Delta m_{21}^{2} &=  7.55^{+0.20}_{-0.16} \times 10^{-5} \,  \text{eV}^{2} \, \\
\Delta m_{32}^{2} &=  (2.423 \pm 0.03)\times  10^{-3} \,  \text{eV}^{2} \, .
\end{align}
In~\cite{gravtheta} it was observed that a potential that respects the flavour symmetry could in principle give a hierarchical vacuum and thus hierarchical masses for the neutrinos, but they did not study the dynamics. 
 In this work, we highlight that the leading renormalizable contribution for the neutrino condensate cannot generate such a breaking pattern. The vacuum configuration of the condensate is constrained by the flavour symmetry (assumed to be respected by the anomaly) in both Dirac and Majorana cases. The leading contribution at low energies comes from the renormalizable potential, which is the same for both Dirac and Majorana neutrinos, although the two possibilities may be distinguished using astrophysical neutrinos~\cite{Funcke:2019grs} or soft topological defects~\cite{Dvali:2021uvk}. Considering the renormalizable potential, there are only two possible symmetry-breaking patterns~\cite{su3min1,su3min2}. The first one is where all the neutrinos acquire the same vacuum and thus the same mass, and the second one is where only one neutrino acquires a non-zero vacuum expectation value and the other two remain massless.

The interaction between the condensate and the gravitational instantons, which generates the low energy potential, could generate higher-dimensional operators which are expected to be subleading at low energies. If the effects of these operators at low energy are not negligible, then the inclusion of higher-dimensional operators could allow hierarchical breaking to occur. Without a deeper understanding of the low energy description of these interactions, it is not clear that gravity could generate the splittings by itself. We therefore also explore the possibility of new physics that contributes to the splitting of the different flavours of neutrinos. 

If the new physics generates a mass contribution for the neutrinos in the UV, this breaks the anomalous symmetry explicitly in the low energy and can spoil the mechanism. The UV contribution could be small, and then the hard mass at low energy behaves similarly to the up and down quark masses in QCD. In this work, we explore the scenario where the anomaly at low energies generates a hierarchical solution, and the contributions from new physics only affect the splittings and do not generate a universal contribution. This scenario is also motivated in order to have a novel cosmological evolution~\cite{Dvali:2017mpy} since the phase transition occurs at very late times, and the neutrinos are truly massless in the early universe. Depending on the time at which the phase transition occurs, we can either have no information about the neutrino mass from cosmology, or the bounds only apply only to the lightest neutrino species.\footnote{If there is a partial contribution from the new physics in the UV, then the cosmological mass limits~\cite{PhysRevLett.105.031301,Riemer-Sorensen:2013jsa, Giusarma:2016phn,Vagnozzi:2017ovm,Giusarma:2018jei,Tanseri:2022zfe} of $\sum_{\nu} m_{\nu} < [0.28,0.14]~\text{eV}$ can be applied to the generated mass.}

We consider two different scenarios where the new physics could appear. The first one is where new physics contributes to neutrino masses in loop processes. We can imagine a scenario where the condensate is universal and the degeneracy is broken by loop contributions of a heavy sector. We show that the Standard Model (SM) $W$-boson generically does generate a splitting, but the splitting is too small to accommodate observed neutrino masses. We discuss how difficult it is to create a predictable contribution for the splitting at the loop level in general UV extensions. 

Another possibility for new physics is with new scalars that interact with the condensate and change the scalar potential. 
We consider two possibilities: the first is for Majorana neutrinos masses where we introduce a $\mathbf{3}$ of the flavour $SU(3)$, and the second is for Dirac neutrinos where we introduce $\mathbf{(1,3)}$ and $\mathbf{(3,1)}$ of the $SU(3)_L \times SU(3)_R$ flavour symmetry. If these new scalars are heavy, they can be integrated out and described by an effective field theory (EFT). This EFT description should also exhibit a hierarchical breaking pattern. We show that the separation of scales is difficult to engineer, which can be understood as the vacuum structure needing to be qualitatively changed by the new physics. 

The remainder of this paper is organized as follows. In section~\ref{sec:phase}, we briefly review the description of the phase transition in terms of the condensate. In section~\ref{sec:neutrinoINT} we construct the interactions of the condensate with the SM particles and calculate the one-loop mass contributions to the neutrinos. In section~\ref{sec:modelbuild} we discuss the problems of model building a predictable loop-level contribution for the neutrino mass splittings. In section~\ref{sec:hiera} we explore one specific UV completion which introduces new interactions to the neutrino condensate. In section~\ref{sec:EFT} we explore the EFT description and how the EFT operators can alleviate the restriction for a hierarchical vacuum. We conclude in section~\ref{sec:conc}.

\section{Description of the phase transition} \label{sec:phase}
Given the SM particle content minimally coupled to gravity, the only assumption necessary for the generation of neutrino masses is that, in pure gravity, we have a non-zero topological vacuum susceptibility at zero momentum,
\begin{align}
\langle R\tilde{R},R\tilde{R} \rangle_{q\rightarrow 0} = \text{const} \neq 0 \, .
\end{align}
Here $R$ is the Riemann tensor, and $\tilde{R}$ is its dual. This is equivalent to the statement that the gravitational $\theta$ angle is physical. Similar to the massless quark solution of the QCD strong $CP$ problem~\cite{anomaly1}, if neutrinos are massless in the absence of gravity, then they will screen the gravitational electric field $E_{G}= R\tilde{R}$ and hence make the gravitational $\theta$-term vanish. If this condition is satisfied,\footnote{At this point we could also have a small hard mass for the neutrinos and the mechanism would still work, similarly to the $\eta'$ in QCD.} then it was shown that the massless fermions $f$ generates a condensate~\cite{gravtheta,Dvali:2017mpy} and the spectrum becomes gapped:
\begin{align}
\Lambda_{G} = \langle R\tilde{R},R\tilde{R} \rangle_{q\rightarrow 0}^{1/8} \sim |\langle \bar{f} f \rangle|^{1/3} \sim  m_{f} \, .
\end{align}
This generates neutrino masses through a Higgs-like composite field. The description of the phase transition at low energy changes depending on the existence of right-handed neutrinos.  If there are no right-handed neutrinos, then the flavour symmetry is $SU(3)$, the neutrino mass is of the Majorana type, and the vacuum is described by:
\begin{align}
\phi_{ij} = \bar{\nu}_{L \, i}^{c} \nu_{L \, j} \, ,
\end{align}
where $\nu_{L}^{c}$ is the charge conjugation state defined as $\nu_{L}^{c} = C \bar{\nu}_{L}^{T}$. $\phi$ transforms as a $\mathbf{6}$ under the flavour symmetry.
If there are right-handed partners of the neutrino, then the flavour symmetry is expanded to $SU(3)\times SU(3)$,\footnote{In principle there do not need to be three right-handed neutrinos, but we assume that is the case for simplicity.}
and the neutrino mass is Dirac. The vacuum is described by:
\begin{align}
\Sigma_{ij} = \bar{\nu}_{i}\nu_{j} \, .
\end{align}
where $\Sigma$ transforms as a bifundamental $(\mathbf{3}, \bar{\mathbf{3}})$ under the flavour symmetry.

The low energy description of the most general potential for both cases is highly similar, so we highlight them simultaneously to understand the possible vacuum configurations. The possible phenomenological signals to probe both cases are studied in~\cite{Funcke:2019grs,Dvali:2021uvk}. The study of a bi-fundamental of $SU(3)\times SU(3)$ is done in~\cite{su3min1,su3min2} and we highlight some of the important results here. The most general renormalizable potential for $\phi$  consistent with the flavour symmetry is given by:
\begin{align} \label{eq:pot}
V(\phi) = -\frac{\mu^{2}_{\phi}}{2} \Tr(\phi^{\dagger}\phi)+ M_{1}\left( \det\phi +\det\phi^{\dagger} \right) + \frac{\lambda_{1}}{4} \Tr(\phi^{\dagger}\phi)^{2} + \frac{\lambda_{2}}{4}\Tr(\phi^{\dagger}\phi\phi^{\dagger}\phi) \, ,
\end{align}
where we can choose $M_{1}$ to be real by using an overall phase redefinition. The potential for the Dirac case is exactly the same given the identification $\phi_{ij} \rightarrow \Sigma_{ij}$. 

Since this potential has only 4 free parameters, it is possible to study all possible minima of the potential. In~\cite{su3min1} it is shown that the breaking of a bi-fundamental representation of $SU(3)_{L} \times SU(3)_{R}$ can only have two distinct patterns. The case of the symmetric representation of $SU(3)$ is the same because the potential is the same. Thus two possible vacua are:
\begin{align}
\Sigma_{ij} = \text{diag}\left(\text{v},\text{v},\text{v} \right),
\end{align} 
or the vev configuration where only one diagonal element gets a non-zero value: 
\begin{align}
\Sigma_{ij} = \text{diag}\left(\text{v},0,0 \right).
\end{align} 
This means that neither the bi-fundamental ($\Sigma_{ij}$) for the Dirac mass nor the symmetric representation ($\phi_{ij}$) for the Majorana mass can generate hierarchical neutrino masses consistent with current experimental data. The explicit calculation that shows the absence of a hierarchical vacuum for the renormalizable potential is done in Appendix \ref{app:noHIERA}. Since the renormalizable potential cannot generate the desired masses for the neutrinos, let us explore how loop corrections to the neutrino mass can change this story.

\section{Low energy interactions for the $SU(3) \rightarrow SO(3)$ breaking} \label{sec:neutrinoINT}
In this section, we investigate the possibility that neutrino interactions generate mass splitting when we start from the configuration where all the vevs are equal.\footnote{ We do not explore the possibility where the initial configuration has two massless neutrinos since it is impossible to lift the degeneracy without generating a hard mass in the UV as we expand further in section~\ref{sec:modelbuild}.}  We focus on the Majorana mass generation, but in the Dirac case, the calculations are similar. The interaction between the neutrino and the condensate excitation field can be described by the following Lagrangian~\cite{gravtheta}:
\begin{align}\label{eq:nuphiLag}
\mathcal{L}_{\phi\nu} = g_{\phi\nu}\left( \phi_{ij}^{\dagger} \bar{\nu}_{L \, i }^{c} \nu_{L \, j} + \phi_{ij} \bar{\nu}_{L \, i} \nu_{L \, j}^{c} \right) 
\end{align}
Below the anomaly scale, the field $\phi_{ij}$ gets a vev generating the neutrino mass, which we assume to be universal:
\begin{align}\label{eq:phivev}
\phi_{ij} = \text{v}_{\phi} \mathbb{I}_{3 \times 3} \, .
\end{align}
The Goldstone modes $\pi_{\hat{a}}$, and the radial modes $\phi_{i}$ and $\eta_{i}$ can be described after the symmetry breaking using the following Lagrangian:\footnote{Notice that the Majorana fermion, $\nu_{M} = \nu_{L} + \nu_{L}^{c}$,   violates the initial $SU(3)$ since we are mixing $3$ and $\bar{3}$ representations. However, we can use this description after the symmetry breaking since the remaining symmetry is $SO(3)$.}
\begin{align}
\mathcal{L}_{\pi\nu} =  \frac{1}{8f_{\phi}} \bar{\nu}_{M \, i} \gamma^{\mu}\gamma^{5}\nu_{M \, j} \partial_{\mu}\pi_{\hat{a}} \lambda^{\hat{a}}_{ij}  + \mathcal{O}(\partial^{2}) \, .
\end{align}
\begin{align} \label{eq:lag}
\mathcal{L}_{\phi\nu} = \left( g_{\phi\nu}v_{\phi} + \frac{g_{\phi\nu}}{\sqrt{2}} \phi_{i}\right)  \bar{\nu}_{M \, i}^{c} \nu_{M \, i }  + \frac{i g_{\phi\nu}}{\sqrt{2}} \eta_{i}\bar{\nu}_{M \, i }^{c}\gamma_{5} \nu_{M \, i} \, .
\end{align}
We write the neutrinos as Majorana states, $\nu_{M} = \nu_{L} + \nu_{L}^{c}$, and $\hat{a}$ runs over broken generators. After the breaking all neutrinos obtain the same mass:
\begin{align}\label{eq:nmass}
m_{\nu \, i } = m_{0} = 2g_{\phi\nu}v_{\phi} \, .
\end{align}

The mass basis that we are working in is not necessarily equal to the charge basis of the Standard Model, we can define the unitary rotation between them as:
\begin{align}
\nu_{L \, \alpha} = \sum_{i} U_{\alpha i} \nu_{L \, i} \, , 
\end{align}
where $\alpha$ runs over the charge basis $e, \mu, \tau$. Then we can write the neutrino interactions from the SM in the mass basis as:
\begin{align}
\mathcal{L}_{SM\nu} = -\frac{g}{4c_{\theta_{w}}} \bar{\nu}_{M \, i} \slashed{Z} \gamma_{5} \nu_{M \, i} + \frac{g}{\sqrt{2}} \left(U_{i \alpha}^{\dagger} \bar{\nu}_{M \, i}\slashed{W}^{+}\mathcal{P}_{L} l_{\alpha}+  U_{\alpha i} \bar{l}_{\alpha} \mathcal{P}_{R} \slashed{W}^{-} \nu_{M \, i} \right) \, .
\end{align}
Now let us study the one-loop contribution for the mass to generate the expected splittings.

\subsection{One-loop mass generation from Standard Model} \label{sec:Wcontribution}
We investigate the one-loop self-energy of the neutrino to see if the Standard Model can account for the small mass splittings between the neutrinos. Because the Goldstone boson ($\pi_{\hat{a}}$) and the radial modes ($\phi_{i}$, $\eta_{i}$), in the absence of extended symmetries, cannot break the symmetry further at loop level~\cite{golds1,golds2,golds3}, they contribute equally to the mass and can be ignored. This can be understood from the Lagrangian in Eq.~\eqref{eq:lag} being diagonal in the mass basis and with a universal coupling. This is also the case for the $Z$ boson since the interaction ends up being diagonal in the mass basis as well. Since the initial mass value is already a free parameter, these shifts can be re-absorbed in the definition of the initial mass. The only non-trivial contribution comes from the $W$-boson interaction. We compute the $W$ exchange self-energy as:
\begin{align}
i \Sigma_{ij}^{W}(p^{2}) = -\frac{g^{2}}{2} \sum_{\alpha} U^{\dagger}_{i\alpha}U_{\alpha j} \int \frac{\dd[4]{k}}{(2\pi)^{4}} \frac{\gamma^{\mu}\mathcal{P}_{L}\left((\slashed{p}-\slashed{k})+m_{l_{\alpha}} \right)\gamma_{\mu}\mathcal{P}_{L}}{\left( (p-k)^{2}-m_{l_{\alpha}}^{2} \right) \left( k^{2}-m_{W}^{2}\right)} \, ,
\end{align}
where only the transverse modes couple to the Majorana states. This integration is straightforward and gives:
\begin{align}
\Sigma_{ij}^{W}(p^{2}) = \slashed{p} \mathcal{P}_{L}\frac{g^{2}}{16\pi^{2}} \sum_{\alpha}  U^{\dagger}_{i\alpha}U_{\alpha j} \left( \mathcal{B}(p^{2};m_{W},m_{l_{\alpha}}) + \mathcal{B}_{1}(p^{2};m_{W},m_{l_{\alpha}})  \right)\, .
\end{align}
Where the $\mathcal{B}$ functions are the Passarino-Veltman coefficients~\cite{Denner:2005nn}. The term proportional to  $\mathcal{P}_{R}$ of the self-energy has the same coefficient because of the Majorana condition. It can be obtained by computing the process with the anti-particle. 

The divergence of this self-energy is diagonal and renormalizes the wave function of the neutrinos. Because the contribution is divergent, the actual value for each mass is not a prediction from the theory. However, since the divergence is universal, differences between the masses are finite and thus a prediction of the theory. For simplicity, we use the $\overline{\text{MS}}$ renormalization scheme. We can then write the contribution for the mass as:
\begin{align}
m_{\nu \, i} = m_{\nu}^{0} +  m_{\nu}^{0}\frac{g^{2}}{16\pi^{2}} \sum_{\alpha}  U^{\dagger}_{i\alpha}U_{\alpha i}\left( \frac{1}{4}- \frac{m_{l_{\alpha}}^{2}}{2m_{W}^{2}}+ \frac{m_{\nu}^{0 \, 2}}{6m_{W}^{2}} - \log \frac{m_{W}}{m_{\nu}^{0}} + \mathcal{O}(1/m_{W}^{4}) \right) \, .
\end{align}
The mass splitting between two different mass states is then given as:
\begin{align}\label{eq:wsplitting}
m_{\nu \, i}^{2} -m_{\nu \, j}^{2} \approx    - 
\left(\frac{m^{0}_{\nu}}{m_{W}} \right)^{2}\frac{g^{2}}{16\pi^{2}}\sum_{\alpha}  \left(  U^{\dagger}_{i\alpha}U_{\alpha i} -  U^{\dagger}_{j\alpha}U_{\alpha j}  \right) m_{l_{\alpha}}^{2} \, .
\end{align}
Using the fact that $m_0 \lesssim 1$ eV and that the unitary matrix elements are bounded $|U_{i\alpha}| \leq 1$, we can put an upper bound on the splitting
\begin{equation}\label{eq:Wsplitt} 
| m_{\nu \, i}^{2} -m_{\nu \, j}^{2} | \lesssim 7\times 10^{-7} \text{eV}^{2} \, ,
\end{equation}
which is far too small to accommodate either measured splitting. The question now is can we introduce new particles to make the splitting consistent with observations?

\section{Model Building the neutrino splittings and UV sensitivity} \label{sec:modelbuild}
Generating neutrino mass splittings at the loop level is a delicate endeavour. The complication arises because we do not want to generate any new universal mass contribution for the neutrinos. Since we do not have chiral symmetry in the Majorana case, it is difficult to avoid such masses. Additionally, we need the mass difference to be finite to predict it in the model, which can be challenging to construct. This is a bigger problem if we try to lift the vacuum with two massless neutrinos. Usually, the splittings are proportional to the difference in the couplings for different flavours, which also contributes to the divergent piece.  

One way to have finite splittings is to mimic the SM $W$ interaction, in which the splittings come from the diagonalization matrix. This, however, does not work for a simple $W'$ extension, since the contribution from $W$ is already orders of magnitude smaller than the experimental value. Another possibility is to include additional generations, which are heavily constrained and the contribution will be suppressed by the mixing matrix.

If we try to generalize this approach, the diagonalization matrix needs to appear in the interaction universally. We need a new set of fermions that interact with the neutrinos. The mediator, in principle, can be a scalar of a vector. However, because of the chirality flip nature of the scalar interaction, there is always a term proportional to the identity in the self-energy. This term is proportional to the fermion mass and thus gives a non-zero contribution to the mass in the neutrino chiral limit. The mass splitting could be UV sensitive and, once the new physics is known, we would be able to resolve this arbitrariness. 

One example of a simple model that generates neutrino splittings in a non-predictable way is a $Z'$ extension of the SM. At low energies we have the following interaction:
\begin{align}
\mathcal{L}_{Z'} = -\frac{g_{\alpha \beta}}{2}\bar{\nu}_{L \, \alpha}\slashed{Z}' \mathcal{P}_{L}\nu_{L \, \beta}  \, .
\end{align}
We can assume for simplicity that the coupling is diagonal in the SM gauge basis and lepton specific: $g_{\alpha \beta}= \text{diag} ( g_{e},g_{\mu}, g_{\tau})$. The same couplings are present in the charged lepton sector where most of the constraints are derived. The UV completion of such a model is responsible for the generation of the $Z'$ mass. In a UV complete model, there will be finite contributions to the neutrino mass. The neutrino mass will then run as a divergent quantity up to the new physics scale and then stop at some finite value. The $Z'$ interaction can be written on the mass basis using Majorana states:
\begin{align}
\mathcal{L}_{Z'} &= -\frac{g_{\alpha \beta}}{2} U^{\dagger}_{i \alpha} U_{\beta j} \bar{\nu}_{M \, i}\slashed{Z}' \gamma_{5}\nu_{M \, j} =  -\frac{g^{U}_{i j}}{2} \bar{\nu}_{M \, i}\slashed{Z}' \gamma_{5}\nu_{M \, j} \, ,
\end{align}
where we define $g^{U}_{ij} = g_{\alpha \beta} U^{\dagger}_{i \alpha} U_{\beta j} $. 
The neutrino self-energy in the mass basis can be written as:

\begin{align}
i \Sigma_{ij}^{Z'}(p^{2}) = - \sum_{k} g^{U \, \dagger}_{ik}g^{U}_{kj} \int \frac{\dd[4]{k}}{(2\pi)^{4}} \frac{\gamma^{\mu}\gamma_{5}\left((\slashed{p}-\slashed{k})+m_{\nu_{k}} \right)\gamma_{\nu}\gamma_{5}}{\left( (p-k)^{2}-m_{\nu_{k}}^{2} \right) \left( k^{2}-m_{Z'}^{2}\right)} \left( -\eta^{\mu \nu} + \frac{k^{\mu} k^{\nu}}{m_{Z'}^{2}} \right) \, ,
\end{align}

We can perform the integral and introduce a cutoff $\Lambda$ where new physics contributes to the neutrino mass. The most important contribution  can be written as:

\begin{align}
\Sigma_{ii}^{Z'}(m_{0}^{2}) = \frac{m_{0}}{16\pi^{2}}\sum_{k} g^{U \, \dagger}_{ik}g^{U}_{ki} \left( \frac{4}{3} + 9 \log\frac{\Lambda^{2}}{m_{Z'}^{2}} \right) \, .
\end{align}
We can then calculate the squared mass difference between two mass eigenstates:
\begin{align}
m_{\nu \, i}^{2} -m_{\nu \, j}^{2} \approx   \frac{m_{0}}{8\pi^{2}}\sum_{k} \left( \frac{4}{3} + 9 \log\frac{\Lambda^{2}}{m_{Z'}^{2}} \right) \left(g^{U \, \dagger}_{ik}g^{U}_{ki} - g^{U \, \dagger}_{jk}g^{U}_{kj} \right)
\end{align}
which is in fact divergent, thus rendering the mass splitting not a prediction of the theory. This same analysis of the unpredictability of the splitting can be done with the vacuum where we start with two massless neutrinos.

This calculation highlights that there are not many results that we can draw from these UV completions in this description if the physics at the cutoff is not known. A more detailed model which has a finite contribution could still exist, and we leave this possibility for further investigation.

\section{ Hierarchical vacuum from new neutrino condensate interactions.} \label{sec:hiera}

In this section, we explore how new physics can modify the condensate interactions such that we have hierarchical vevs. We first assume that the leading description of the condensate comes from the renormalizable couplings to new fields. In section~\ref{sec:EFT}, we also explore the possibility of significant contributions from non-renormalizable interactions.

One example of a model which can generate hierarchical vevs is presented for the Dirac case in~\cite{su3min2}. In this model, we add two fields transforming under the $SU(3)_L\times SU(3)_R$ flavour group as:
\begin{align}
Z_{L} \in (\textbf{3},\textbf{1}) \, , \, Z_{R} \in (\textbf{1},\textbf{3}) \, . 
\end{align}
Using the fact that $\Sigma$ transforms as a $(\textbf{3},\bar{\textbf{3}})$, the most general renormalizable potential can be written as:
\begin{align} \label{eq:DI}
V_{\text{Dirac}}(\Sigma,Z_{L},Z_{R}) = V_{0}(\Sigma) &+ \mu_{L}^{2} Z_{L}^{\dagger}Z_{L} + \mu_{R}^{2} Z_{R}^{\dagger}Z_{R}+ \lambda_{LL} (Z_{L}^{\dagger}Z_{L})^{2} + \lambda_{RR} (Z_{R}^{\dagger}Z_{R})^{2} \, \\ \nonumber
 &+ \lambda_{LR} (Z_{L}^{\dagger}Z_{L})(Z_{R}^{\dagger}Z_{R})  + \lambda_{TLL} \Tr(\Sigma^{\dagger}\Sigma) Z_{L}^{\dagger}Z_{L} + \lambda_{TRR} \Tr(\Sigma^{\dagger}\Sigma) Z_{R}^{\dagger}Z_{R} \\ \nonumber
 &+ \lambda_{L\Sigma\Sigma L} Z_{L}^{\dagger} \Sigma \Sigma^{\dagger} Z_{L} +  \lambda_{R\Sigma\Sigma R} Z_{R}^{\dagger} \Sigma^{\dagger} \Sigma Z_{R} + \left(A_{L\Sigma R} Z_{L}^{\dagger}\Sigma Z_{R} + \text{h.c}\right) \\ \nonumber
 &+ (\lambda_{\epsilon}\epsilon_{ijk}\epsilon_{abc}\Sigma_{ia}\Sigma_{jb}Z_{L}^{k}Z_{R}^{c \, \dagger} + \text{h.c}) \, ,
\end{align}
where $V_0$ is given in Eq.~\eqref{eq:pot} with $\phi \rightarrow \Sigma$. The potential $V_{\text{Dirac}}$ has one additional invariant operator that was not included in~\cite{su3min2}. However, the overall analysis done in the work remains qualitatively similar and is only slightly modified when $\lambda_{\epsilon}$ is non-zero.

The Majorana case is similar. The flavour group  is $SU(3)$, the condensate field $\phi$ is a $\mathbf{6}$, and we add just one fundamental field:
\begin{align}
Z_{M} \in \textbf{3} \, .
\end{align}
The potential is then
\begin{align}\label{eq:MA}
V_{\text{Majorana}}(\phi,Z_{M}) = V_{0}(\phi) &+ \mu_{M}Z_{M}^{\dagger}Z_{M} + \lambda_{MM}(Z_{M}^{\dagger}Z_{M})^{2} + \lambda_{TMM} \mathcal{T}Z_{M}^{\dagger}Z_{M}  \\ \nonumber
&+ \lambda_{M\phi\phi M} Z_{M}^{\dagger}\phi^{\dagger}\phi Z_{M} +\left(A_{M\phi M} Z_{M}^{T}\phi Z_{M} + \text{h.c}\right) \\\nonumber
& +(\lambda_{\epsilon}\epsilon_{ijk}\epsilon_{abc}\phi_{ia}\phi_{jb}Z_{M}^{k}Z_{M}^{c} + \text{h.c}) \, .
\end{align}

Because these potentials are much more complicated than the version without additional fields, obtaining analytical results becomes difficult. We, therefore, explore these models numerically to verify that such potentials can indeed generate hierarchical configurations consistent with observation. We analyze the Majorana case in-depth because it has fewer free parameters, but the overall behaviour of the Dirac potential should be similar. Additionally, from the scans, we obtain that both models can generate the normal and inverted hierarchy. We generated solutions where the lightest neutrino is both massive and massless. 

To apply the experimental constraints of the neutrino masses we need to set a scale $\text{v}_\phi$ and choose a coupling $g_{\phi \nu}$ (see Eqs.~\eqref{eq:nuphiLag} and~\eqref{eq:phivev}). We choose $g_{\phi\nu} = 1/2$ for simplicity such that each neutrino mass can be identified directly as the vev value from Eq.~\eqref{eq:nmass}. Because there are different mass parameters in the theory we re-scale everything such that we have the largest possible neutrino mass which satisfies current experimental bounds~\cite{Super-Kamiokande:2017yvm,PDG,KATRIN:2021uub}:
\begin{align}
 m_{\nu} &\leq 0.8 \,  \text{eV} \, ,\\
\Delta m_{21}^{2} &=  7.55^{+0.20}_{-0.16} \times 10^{-5} \,  \text{eV}^{2} \, \\
\Delta m_{32}^{2} &=  (2.423 \pm 0.03)\times  10^{-3} \,  \text{eV}^{2} \, .
\end{align}

\begin{figure}[t] 
 \resizebox{0.48\linewidth}{!}{\includegraphics[]{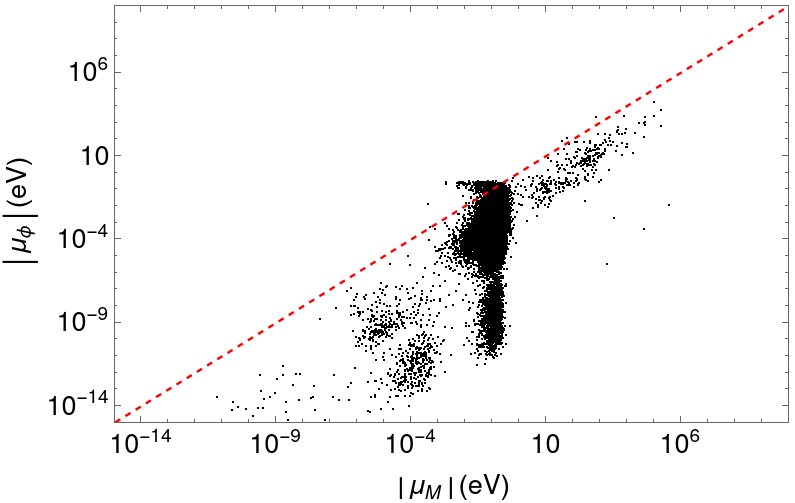}}
  \resizebox{0.48\linewidth}{!}{\includegraphics[]{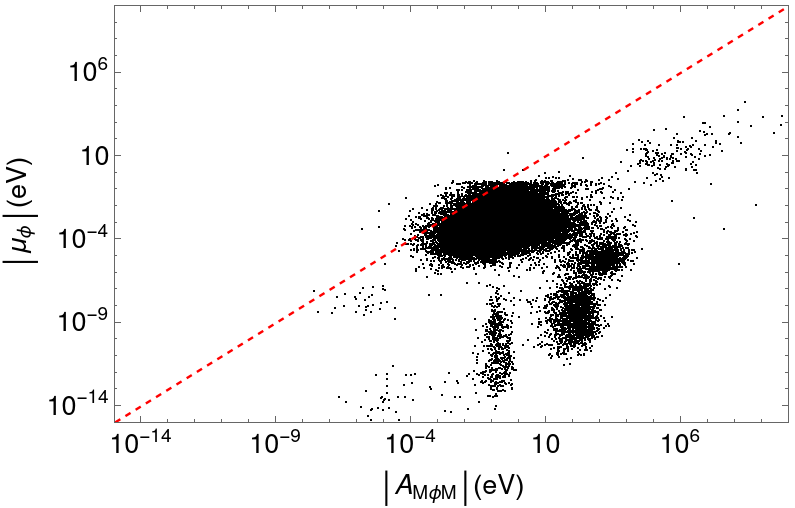}}
\caption{Numerical scans for the Majorana model with $Z_{M}$. All points agree with current neutrino mass observations. We explore different regions of relations between the scale of $\phi$ controlled by $\mu_{\phi}$ and the scale of the new sector $\mu_{M}$ and $A_{M\phi M}$. The dashed red line is where the two mass parameters are equal. \label{fig:fig1}}
\end{figure}

We can now explore whether we have any allowed points where the new fields $Z$ are parametrically heavier than the $\phi$ fields in order to explore a possible EFT description. 
We implement different separations of scales for the scans in order to populate the parameter space in the important regions where we want these new particles to decouple while having hierarchical neutrino masses. In the new sector, we have two massive couplings that control the possible separation of scales from the neutrino condensate. It is expected that when $\mu_{M}$ and $A_{M\phi M}$ are large compared with $M_{1}$ and $\mu_{\phi}$ that we can get some separation between the sectors. In Figure~\ref{fig:fig1} we compare the dimensionful Lagrangian parameters of the new sector to those of the $\phi$ for points that satisfy all experimental observations. We see that for many of the points the scales are comparable, but there are also points with large separations.
 
\begin{figure}[b] 
  \resizebox{0.48\linewidth}{!}{\includegraphics[]{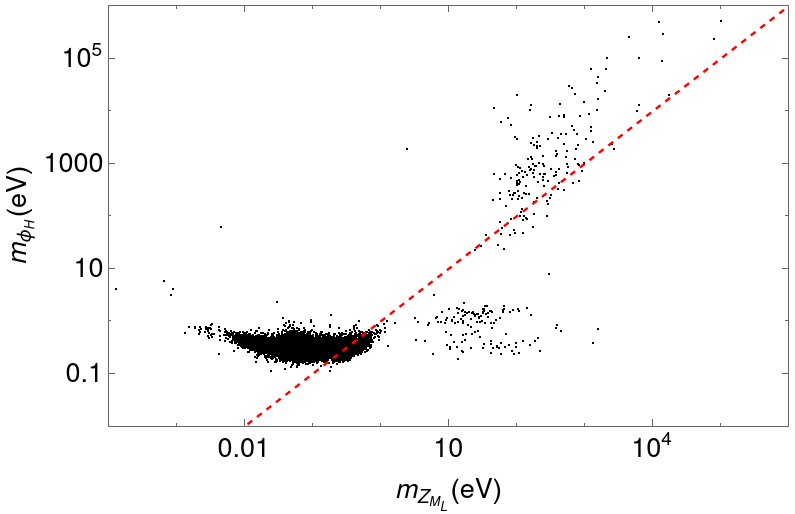}}
    \resizebox{0.48\linewidth}{!}{\includegraphics[]{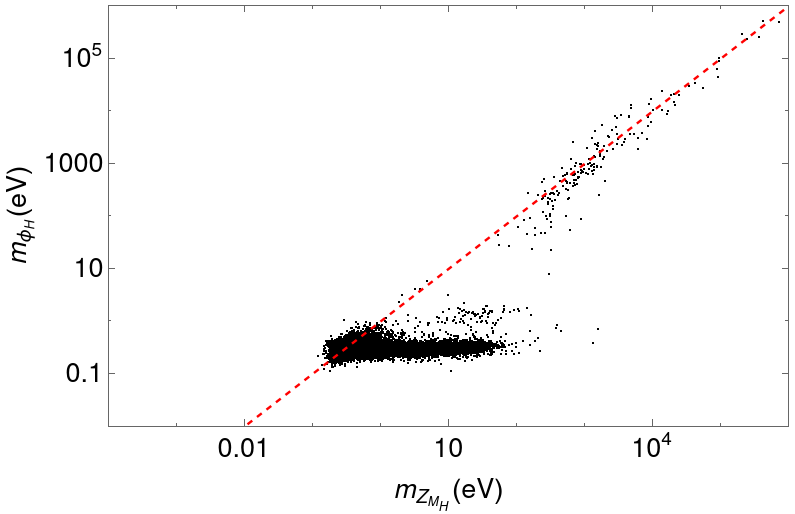}}
\caption{Numerical scans for the mass of lightest (left) and heaviest (right) new sector particle $Z_{M_{L}}$/$Z_{M_{H}}$ vs.~the mass of the heaviest condensate particle $m_{\phi_{H}}$ for the Majorana model with $Z_{M}$. All points satisfy current neutrino mass observations. 
The dashed red line is where the two masses are equal. \label{fig:fig2}}
\end{figure}

 The story is slightly different if we look at the spectrum of states rather than Lagrangian parameters. We can say that the separation of scales works if we can separate the massive $\phi$ fields from the massive $Z_{M}$ fields.  In Figure~\ref{fig:fig2} we show the relation between the heaviest $\phi$ state and the lightest (left) or heaviest (right) $Z_M$ state. We can see that there is a correlation between them. When the mass of the $Z_{M}$ grows, the $\phi$ mass generically does as well, making it difficult to have a  large separation of scales, even if the massive couplings themselves can have significant separations. Note that all the new states are light enough that they are kinematically accessible to collider experiments, but we leave the exploration of its phenomenology to future work. 

We do have some points where there is a significant separation of scales both in the Lagrangian and the spectrum, but as we will see in the next section, the description using effective field theory does not work even for those extreme points. 
This is to be expected in any model where new physics modifies the vacuum structure and at the same time attempts to decouple from the low energy.

\subsection{Cosmological Constraints}

When considering the additional light fields $Z_{M}$ or $Z_{L/R}$ there is the possibility of modifying the evolution of our universe such that the model is inconsistent with observations. This can even be a problem for the condensate $\phi$ if the transition happens before $T \sim 256$ meV. Above these temperatures, Majoron-like couplings are strongly restricted~\cite{Hannestad:2005ex,Basboll:2008fx,Archidiacono:2013dua}. We are here assuming that the transition occurs below this temperature, and thus for these cosmological bounds, the condensate $\phi$ does not exist and there are only free neutrinos. This is not the case for any additional sector which we include, these fields will be present in the early universe and have a portal to the SM through the neutrinos. We can expect, from Eq.~\eqref{eq:DI} and Eq.~\eqref{eq:MA} that before the phase transition there is a non-renormalizable interaction between the neutrinos and these new fields of the form:
\begin{align}
\mathcal{L}_{\text{int}}^{M} = \frac{A_{M\phi M}}{f_{\phi}^{2}} Z_{M}^{\dagger}\bar{\nu}_{L }^{c} \nu_{L } Z_{M} \, , \nonumber\\
\mathcal{L}_{\text{int}}^{D} = \frac{A_{L\Sigma R}}{f_{\phi}^{2}} Z_{L}^{\dagger}\bar{\nu} \nu Z_{R} \, ,
\label{eq:nuOperator}
\end{align}
for the Majorana and Dirac cases respectively. There are also terms with more neutrino fields suppressed by higher powers of $f$. We can then check if these particles are in thermal equilibrium with the neutrinos when the neutrinos have a temperature $T \gg m_\nu$. If we want this new sector to not be in thermal equilibrium we need to satisfy the following condition:
\begin{align}
n_{\nu} \sigma_{\nu \nu \rightarrow ZZ} \text{v}_{\nu} \lesssim \frac{T^{2}}{M_{P}} \, ,
\end{align}
where $M_{P}$ is the Planck mass. The neutrinos are in thermal equilibrium and relativistic so we have that $\text{v}_{\nu} \sim 1$ and $n_{\nu} \sim T^{3}$. We can re-write the condition as:
\begin{align}
\sigma_{\nu \nu \rightarrow ZZ}\lesssim \frac{1}{ T M_{P}} \, .
\end{align}
The process $\nu \nu \rightarrow ZZ$ is mediated at tree level by the operators in Eqs.~\eqref{eq:nuOperator}. The cross section at leading order is constant,\footnote{A similar cross-section occurs for the dimension 5 Weinberg operator in the SM~\cite{Fuks:2020zbm}.} and we can write for both cases as:
\begin{align}
\sigma_{\nu \nu \rightarrow ZZ} \sim \frac{A^{2}}{f_{\phi}^{4}} \, .
\end{align}
We can go one step beyond and use the fact that the trilinear constant $A$ is around the same scale as the mass of the new fields because of stability. This fact is also evident from the right panel of Fig.~\ref{fig:fig1}. We can therefore write the condition for both cases as:
\begin{align}
\frac{m_{Z_{M/L/R}}^{2}}{f_{\phi}^{4}} \lesssim \frac{1}{ T M_{P}}  \, .
\label{eq:bound}
\end{align}
This relation is just for order of magnitude estimates, and since we have the unknown non-perturbative coefficient $f_{\phi}$ this is the best that we can do.

From Eq.~\eqref{eq:bound}, we see that the bound is most stringent at the highest temperature that the neutrinos achieve in the early universe. We can take a relatively low reheating temperature of $T_\text{RH} \sim$ GeV and find the following limit on the mass and $f_\phi$ parameters:
\begin{align}
\frac{m_{Z_{M/L/R}}}{f^{2}_{\phi}} \lesssim 10^{-19} \frac{1}{\text{eV}} \left( \frac{1 \, \text{GeV}}{T_\text{RH}} \right) \, .
\end{align}
We can now speculate what is the size of $f_{\phi}$ in order to have a estimate for the bound on the masses. Since this is an anomaly driven process, we can expect that this interaction only takes place close to the condensate formation, and since this is gravity driven we can expect that $f_{\phi} \sim M_{P}$. If this is the case, then we have a suppressed interaction with the neutrinos and a trivial bound on the mass:
\begin{align}
m_{Z_{M/L/R}} \lesssim 10^{37} \text{eV}  \left( \frac{1 \, \text{GeV}}{T_\text{RH}} \right)\, .
\end{align}
If, on the other hand, we assume that the condensate constant is around the condensate scale, $f_{\phi} \sim \text{eV}$, then we can find a very strong bound on the mass of these new states:
\begin{align}
m_{Z_{M/L/R}} \lesssim 10^{-16} \text{eV}\left( \frac{1 \, \text{GeV}}{T_\text{RH}} \right) \, .
\end{align}
In this case, because the scale is so low, other higher dimensional operators suppressed by higher powers of $f_\phi$ will also contribute on equal footing, making the effective field theory expansion not converge. This problem appears also when we integrate out this new sector trying to separate from $\phi$, as we show in the next section. For both estimations of $f_{\phi}$ it is possible to still generate the correct neutrino mass hierarchies. In the case where $f_{\phi}$ is close the the condensate scale there significant fine-tuning involved.

Another possibility is that rather than rarely or never producing these new fields in the early universe, is that they are sufficiently heavy so as to not be relativistic at the time of Big Bang Nucleosynthesis. In this case they would contribute to dark matter, and one must also ensure that they do not overclose the universe. Further, for them to not be too warm, they would need to have a mass above 6.5 keV~\cite{DES:2020fxi} if they saturate the dark matter relic density. This would also require fine-tuning to get the right neutrino mass scale, and further exploration of this possibility is outside our scope. 

Additional model-independent cosmological bounds for late-time neutrino generation can be found in~\cite{Lorenz:2018fzb,Franca:2009xp,PhysRevD.106.103026,la2013mass}. Overall, it is challenging to determine how restrictive late-cosmology bounds are for these models without a more profound understanding of the condensation mechanism and how to estimate $f_{\phi}$.

\section{Effective Field theory description of the condensate}\label{sec:EFT}

In this section, we study the possibility of deforming the renormalizable potential with higher-dimensional operators to generate a hierarchical vacuum. These higher-dimensional operators could come from the gravitational instanton interactions. We also explore if the UV complete model in the previous section can have a low energy description in the EFT language. 

The description of the potential is the same for both Majorana and Dirac neutrinos, given the identification $\phi_{ij} \rightarrow \Sigma_{ij}$. The analysis for the Dirac case is similar to the work~\cite{su3min2}, where they study the hierarchical pattern for the quark sector. In exploring higher dimensional operators, it is more convenient to use the following flavour invariant operators: 
\begin{align}
\mathcal{T} = \Tr(\phi^{\dagger}\phi) \, , \, \mathcal{A} = \Tr(Adj(\phi^{\dagger}\phi)) \, , \, \mathcal{D}^{2} = \det(\phi^{\dagger}\phi) \, .
\end{align}
The adjugate of a matrix is defined as $Adj(M)=\det(M)M^{-1}$. In addition, because of the special\footnote{Namely the symmetry group is $SU(3)$, not $U(3)$.} nature of the symmetry group, we have also that $\mathcal{D}=\det\phi$ is invariant. Every operator that is invariant under the symmetry group can be written as powers of these three operators~\cite{su3min2}. We can then describe any hierarchical configuration $(\text{v}_{1},\text{v}_{2},\text{v}_{3})$ in terms of the invariants $(\mathcal{T},\mathcal{A},\mathcal{D})$. 

Now that we can describe the vacuum in terms of the invariants, we can try to generate a configuration which generates a hierarchical solution. We can write the full list of higher-dimensional operators up to dimension 8:
\begin{align}\label{eq:EFF1}
V_{5} &= \frac{c_{5}}{\Lambda}\mathcal{T}\mathcal{D} \, , \\ \label{eq:EFF2}
V_{6} &= \frac{c_{6}^{(1)}}{\Lambda^{2}} \mathcal{T}^{3} + \frac{c_{6}^{(2)}}{\Lambda^{2}} \mathcal{A}\mathcal{T} + \frac{c_{6}^{(3)}}{\Lambda^{2}} \mathcal{D}^{2} \, , \\ \label{eq:EFF3}
V_{7} &= \frac{c_{7}^{(1)}}{\Lambda^{3}} \mathcal{T}^{2}\mathcal{D} + \frac{c_{7}^{(2)}}{\Lambda^{3}} \mathcal{A}\mathcal{D} \, , \\  \label{eq:EFF4}
V_{8} &= \frac{c_{8}^{(1)}}{\Lambda^{4}} \mathcal{T}^{4} +   \frac{c_{8}^{(2)}}{\Lambda^{4}} \mathcal{T}^{2} \mathcal{A} + \frac{c_{8}^{(3)}}{\Lambda^{4}} \mathcal{A}^{2} + \frac{c_{8}^{(4)}}{\Lambda^{4}} \mathcal{T}\mathcal{D}^{2} \, .
\end{align}
As a simple exercise, we begin by turning on only the following two operators:
\begin{align}
V &= V_{0}+ \frac{c_{6}^{(3)}}{\Lambda^{2}} \mathcal{D}^{2}   + \frac{c_{8}^{(3)}}{\Lambda^{4}} \mathcal{A}^{2} \, , \\
V_{0} &= -\frac{\mu^{2}_{\phi}}{2} \mathcal{T} + \lambda_{T} \mathcal{T}^{2} + \lambda_{A} \mathcal{A} - 2 M_{1} \mathcal{D} \, ,
\end{align}
where $V_0$ is the potential from Eq.~\eqref{eq:pot}, and we identify $\lambda_{T} =\frac{\lambda_{1} + \lambda_{2}}{4}$ and $\lambda_{A} = -\frac{\lambda_{2}}{2}$. The minimization condition in the $(\mathcal{T},\mathcal{A},\mathcal{D})$ field basis for the potential is:
\begin{align}
\frac{\partial V}{\partial \mathcal{T}} &= - \frac{\mu_{\phi}^{2}}{2} + 2\lambda_{T}\mathcal{T}  = 0 \, \\
\frac{\partial V}{\partial \mathcal{A}} &= \lambda_{A} + 2\frac{c_{8}^{(3)}}{\Lambda^{4}} \mathcal{A}  =  0 \, ,  \\
\frac{\partial V}{\partial \mathcal{D}} &= -2M_{1} + 2\frac{c_{6}^{(3)}}{\Lambda^{2}} \mathcal{D}  =  0 \, .  
\end{align}
The second derivative matrix in the $(\mathcal{T},\mathcal{A},\mathcal{D})$ basis can also be written to investigate if the solution is a minimum:
 \begin{align}
 \mathcal{M}^{2} = \begin{pmatrix}
2\lambda_{T} & 0 & 0 \\
0 & 2 \frac{c_{8}^{(3)}}{\Lambda^{4}} & 0 \\
0 & 0 &  2 \frac{c_{6}^{(3)}}{\Lambda^{2}}
\end{pmatrix} \, .
 \end{align}
This indicates that the two Wilson coefficients need to be positive. In this simple example, we can dial the tree-level parameters to generate any arbitrary vacuum parameterized by $(\mathcal{T}_{0}, \mathcal{A}_{0}, \mathcal{D}_{0} )$:
\begin{align}\label{eq:eftsol}
\mu_{\phi}^{2} = 4\lambda_{T} \mathcal{T}_{0} \, , \, \lambda_{A} = - 2\frac{c_{8}^{(3)}}{\Lambda^{4}}\mathcal{A}_{0} \, , \, M_{1} =  2\frac{c_{6}^{(3)}}{\Lambda^{2}}\mathcal{D}_{0} \, .
\end{align}
In order for the EFT to be valid, we need the dimensionful Lagrangian parameters $M_1$ and $\mu$ and also the vevs $\mathcal{T}^{1/2}$, $\mathcal{D}^{1/3}$, and $\mathcal{A}^{1/4}$, to all be parametrically smaller than $\Lambda$. Further assuming that the Wilson coefficients are not unnaturally large $|c| \lesssim 1$, then tunings are required to satisfy Eq.~\eqref{eq:eftsol}, $\lambda_A \ll 1$ and $M_1 \ll \mathcal{D}^{1/3}$. This suggests that if the anomaly induces higher dimensional operators, it is necessary to tune the parameters to generate a hierarchical solution. Note also that the vacuum is controlled by operators of different mass dimensions, indicating poor convergence of the EFT expansion. 

If the EFT descends from a particular UV completion, there will in general be correlations between different Wilson coefficients. We now explore the EFT that descends from the UV complete models described in section~\ref{sec:hiera}.
The leading one-loop matching coefficients up to dimension 8 are written in Appendix~\ref{app:EFTmajo} for the Majorana model, and we have also computed the matching in the Dirac case. There are some interesting results from the matching. First, there is a clear way to differentiate the models: the Dirac case does not generate the dimension 5 or dimension 7 operators at one loop while the Majorana model does. We can also see the role of having the dimensionful coupling $A_{M\phi M}$ (or $A_{L\Sigma R} $ in the Dirac case), where it gives a positive contribution to the coefficient of the $\mathcal{A}^{2}$ operator, while the leading contribution is negative. In the Majorana case this can be seen in:
\begin{align}
   \frac{c_{8}^{(3)}}{\Lambda^{4}} &=+\frac{\left(A_{M \phi M}\right){}^2
   \left(\lambda _{M \phi  \phi  M}\right){}^3}{12 \pi ^2 \left(m_{Z_M}\right){}^6}-\frac{2 \left|\lambda _{\epsilon }\right|{}^4}{3 \pi ^2
   \left(m_{Z_M}\right){}^4}  -\frac{\left(\lambda _{M \phi  \phi  M}\right){}^4}{192 \pi ^2 \left(m_{Z_M}\right){}^4} \, , 
\end{align}

Having this coefficient positive makes it easier to generate a stable global minimum in the $\mathcal{A}$ direction. It is also possible to see a potential danger when $A_{M\phi M}$ is of the same order as the mass of the new states. This would break the convergence of the EFT series in $1/m$ and highlight again the non-decoupling effect of the two sectors. From the EFT matching, we can also see that some couplings are more important than others to guarantee that we have both $\mathcal{D}^{2}$ and $\mathcal{A}^{2}$ operators at low energy. We can set $\lambda_{\epsilon} = \lambda_{TMM} = 0$ and have all the couplings real and still be able to generate both the hierarchical minima and these two operators at low energy.

Finally, we examined some of the specific parameter points of the UV complete model that have a significant separation between the new states and those of the condensate. Performing the one-loop matching up to dimension 8 and then minimizing the EFT potential does not give the same vacuum as the full theory and is unable to generate neutrino masses consistent with observation. This further signals the breakdown of EFT convergence.

A further in-depth investigation of these models is only truly motivated in the event that this mechanism shows to be the one that generates the neutrino masses. There are questions on the matching, and having to flow from the UV to IR with the parameters of the models. The phenomenology of this model can also be interesting in the same context since there are new states that could be accessible to various experiments. The leading particles that would contribute would still be the $\phi$ and the Goldstone bosons since we are creating a hierarchy of scales, which means that the search for this mechanism can be somewhat model independent.

\section{Conclusion} \label{sec:conc}

In this work, we explore the possibility that the neutrino mass is generated at low energies from the anomaly on the gravitational $\theta$ term. This low energy phase transition could be used in conjunction with~\cite{CCphase} to move the cosmological constant from negative to the small positive value we see today.

The description of the phase transition in terms of the neutrino condensate constrains the possible breaking patterns that we can observe. In the initial proposal for this mechanism, it was assumed that the condensate acquires the hierarchical pattern. We show that such a pattern is not possible with the leading renormalizable potential. We also compute the SM loop contributions to the neutrino mass splitting. The only non-zero contribution to the splitting is from the loop of a $W$ and charged lepton, but that contribution is proportional to $m_l^2/m_W^2$ (see Eq.~\eqref{eq:wsplitting}) and too small to accommodate the observed values. 

We then analyze the scenario where the gravitational anomaly generates a universal neutrino mass and new physics generates the splittings.
The new physics could appear in two different descriptions. It could enter as new scalars interacting with the condensate field, or it could enter as new neutrino interactions in the self-energy process. 
We show, however, that new interactions that generate finite splittings are difficult to construct. We highlight one simple model which has a divergent contribution to the splitting, meaning that neither the mass nor the mass differences are a prediction of the low energy theory.

We also explore how new interactions of the condensate can generate a hierarchical vacuum solution. We investigate a specific model that introduces a minimal content of new particles that can generate the experimentally measured neutrino splittings. We explore the parameter space of this model and calculate cosmological limits considering the impact of the new fields on late-cosmology observables. However, we have found that without a more comprehensive understanding of the condensate mechanism, it is not possible to draw any conclusions regarding the exclusion of this matter content. We then show that the separation of scales is difficult to construct in this specific model and discuss that this is expected to be a general feature since we are trying to generate low-energy effects from high-energy physics. We then examine the EFT description of the condensate where higher dimensional operators can be from the specific UV completion or possibly from the gravitational instantons. We show that in the EFT language it is necessary to have cancellations between different orders of the expansion signalling the same problem that we face in the UV complete model. This means that, if this specific UV complete model is the one that generates the neutrino mass splittings in our universe, then we should expect new particles in reach of the current colliders.

The proposed solutions that we create do not exhaust all the possible UV completions or different ways that new physics can contribute to the splittings given that a low energy phase transition generates universal masses. Further investigation into this problem could clarify what signals to look for to test this proposed neutrino mass mechanism. Additionally, further investigation on the specific low energy description of the gravitational instanton interaction with the condensate could shed some light on what is the exact symmetry breaking which occurs, and if we are required to add new physics to describe the neutrino phenomenology.

\section*{Acknowledgments}
\noindent We would like to thank Philip Tanedo and Andr\'e de Gouv\^ea for their helpful discussions. C.H.dL.~and D.S.~are supported in part
by the Natural Sciences and Engineering Research Council of Canada (NSERC). 

\appendix

\section{Absence of hierarchical vacuum for the renormalizable potential} \label{app:noHIERA}
Let us write the most general vacuum configuration for Majorana and Dirac cases, calling the field X. We can use the flavour symmetry in both cases to get rid of the off-diagonal components and two diagonal phases. The last phase can be made global, and thus the most general configuration vev is:
\begin{align}
X = \text{diag}\left( \text{v}_{1}, \text{v}_{2}, \text{v}_{3} \right)e^{i\alpha/3} \, . 
\end{align}
The minimization condition for the potential in Eq.~\eqref{eq:pot}
can be cast in the following form:
\begin{align}\label{eq:M1sin}
0 &= M_{1}\text{v}_{1}\text{v}_{2}\text{v}_{3}\sin\alpha \, , \\
0 &= (\text{v}_{1}-\text{v}_{2}) \left(\text{v}_{3}^{2}\lambda_{1}+\text{v}_{1}\text{v}_{2}\lambda_{2}+\text{v}_{1}^{2}(\lambda_{1}+\lambda_{2})+\text{v}_{2}^{2}(\lambda_{1}+\lambda_{2})-\mu_{\phi}^{2}+2M_{1}\text{v}_{3}\cos \alpha \right) \, , \\
0 &= (\text{v}_{2}-\text{v}_{3})  \left(\text{v}_{1}^{2}\lambda_{1}+\text{v}_{2}\text{v}_{3}\lambda_{2}+\text{v}_{2}^{2}(\lambda_{1}+\lambda_{2})+\text{v}_{3}^{2}(\lambda_{1}+\lambda_{2})-\mu_{\phi}^{2}+2M_{1}\text{v}_{1}\cos \alpha \right) \, , \\
0 &= (\text{v}_{2}+\text{v}_{3})  \left(\text{v}_{1}^{2}\lambda_{1}-\text{v}_{2}\text{v}_{3}\lambda_{2}+\text{v}_{2}^{2}(\lambda_{1}+\lambda_{2})+\text{v}_{3}^{2}(\lambda_{1}+\lambda_{2})-\mu_{\phi}^{2}-2M_{1}\text{v}_{1}\cos \alpha \right) \, .
\end{align}
We will now show that there cannot be a vacuum with three different non-zero values for the $v_i$.
From the first condition, we can see that, given we are assuming that all vevs are different and non-zero, we need to have $M_{1} = 0$ or $\sin\alpha =0$. When $M_{1} = 0$ the angle becomes a flat direction and we can choose any value, while when $M_{1}\neq 0$ the minimization occurs for $\alpha = 0$. We can then set $\alpha = 0$ in general.\footnote{We could also set $\alpha = \pi$ and this would only change $M_{1}$ to $-M_{1}$ and all the results remain unchanged.}

Assuming that we have all vevs different, we can solve the remaining equations. However, this ends up giving us a contradiction, since the only solutions up to permutations are:
\begin{align}
\text{v}_{1} &= \text{v}_{3} = \sqrt{\frac{\mu_{\phi}^{2}\lambda_{2}^{2}-4M_{1}^{2}(\lambda_{1}+\lambda_{2})}{\lambda_{2}^{2}(2\lambda_{1}+\lambda_{2})}} \, , \, \text{v}_{2} = 2\frac{M_{1}}{\lambda_{2}}  \, , \\
\text{v}_{1} &= \text{v}_{3} = 0 \, , \, \text{v}_{2} = \sqrt{\frac{\mu_{\phi}^{2}}{\lambda_{1}+\lambda_{2}}}  \, .
\end{align}
We can see that we have at least two vevs equal or two vevs equal to zero, which implies that we have no solution that satisfies the condition where all vevs are different and thus we prove by contradiction that we cannot have hierarchical vevs for the renormalizable potential.

\section{EFT matching for the Majorana case}
\label{app:EFTmajo}
We perform the matching at one loop up to the leading contribution in $1/m_{Z_{M}}$. The coefficients are matched in the basis of Eq.\eqref{eq:EFF1}-Eq.\eqref{eq:EFF4}:

\begin{align}
 \frac{c_{5}}{\Lambda} &= \frac{\lambda _{\epsilon }  A_{M \phi  M} \left(\lambda _{M \phi  \phi  M}+3 \lambda _{T \text{MM}}\right)}{4 \pi ^2
   m_{Z_M}^2} \, ,
\end{align} 
\begin{align}
 \frac{c_{6}^{(1)}}{\Lambda^{2}} &= \frac{\left(\lambda _{M \phi  \phi  M}\right){}^3+3 \left(\lambda _{M \phi  \phi  M}\right){}^2 \lambda _{T \text{MM}}+3 \lambda _{M \phi 
   \phi  M} \left(\lambda _{T \text{MM}}\right){}^2+3 \left(\lambda _{T \text{MM}}\right){}^3}{96 \pi ^2 m_{Z_M}^2} \, , \\
 \frac{c_{6}^{(2)}}{\Lambda^{2}} &=-\frac{\left(\lambda _{M \phi  \phi  M}\right){}^3+2 \left(\lambda _{M \phi  \phi  M}\right){}^2 \lambda _{T \text{MM}}-16 \left|\lambda
   _{\epsilon }\right|{}^2 \lambda _{T \text{MM}}}{32 \pi ^2 m_{Z_M}^2} \, \\
 \frac{c_{6}^{(3)}}{\Lambda^{2}} &=   \frac{\left(\lambda _{M \phi  \phi  M}\right){}^3+48 \left|\lambda _{\epsilon }\right|{}^2 \lambda _{M \phi  \phi  M}}{32 \pi ^2 m_{Z_M}^2} \, , 
\end{align}
\begin{align}
 \frac{c_{7}^{(1)}}{\Lambda^{3}} &= -\frac{\lambda _{\epsilon } A_{M \phi  M} \left(\left(\lambda _{M \phi  \phi  M}\right){}^2+2 \lambda _{M \phi  \phi  M} \lambda _{T
   \text{MM}}+3 \left(\lambda _{T \text{MM}}\right){}^2\right)}{8 \pi ^2 m_{Z_M}^4} \, , \\
 \frac{c_{7}^{(2)}}{\Lambda^{3}} &=   \frac{\lambda _{\epsilon } A_{M \phi  M} \left(3 \left(\lambda _{M \phi  \phi  M}\right){}^2-8 \left|\lambda _{\epsilon
   }\right|{}^2\right)}{12 \pi ^2 m_{Z_M}^4} \, ,
\end{align}
\begin{align}
 \frac{c_{8}^{(1)}}{\Lambda^{4}} &= -\frac{\left(\lambda _{M \phi  \phi  M}\right){}^4+4 \left(\lambda _{M \phi  \phi  M}\right){}^3 \lambda _{T \text{MM}}+6 \left(\lambda _{M
   \phi  \phi  M}\right){}^2 \left(\lambda _{T \text{MM}}\right){}^2+4 \lambda _{M \phi  \phi  M} \left(\lambda _{T \text{MM}}\right){}^3+3
   \left(\lambda _{T \text{MM}}\right){}^4}{384 \pi ^2 m_{Z_M}^4} \, , \\
\frac{c_{8}^{(2)}}{\Lambda^{4}} &=  \frac{\left(\lambda _{M \phi  \phi  M}\right){}^4+3 \left(\lambda _{M \phi  \phi  M}\right){}^3 \lambda _{T \text{MM}}+3 \left(\lambda _{M
   \phi  \phi  M}\right){}^2 \left(\lambda _{T \text{MM}}\right){}^2-24 \left|\lambda _{\epsilon }\right|{}^2 \left(\lambda _{T
   \text{MM}}\right){}^2}{96 \pi ^2 m_{Z_M}^4} \, ,\\ 
\frac{c_{8}^{(3)}}{\Lambda^{4}} &=   -\frac{\left(\lambda _{M \phi  \phi  M}\right){}^4+128 \left|\lambda _{\epsilon }\right|{}^4}{192 \pi ^2 \left(m_{Z_M}\right){}^4} \, , \\
\frac{c_{8}^{(4)}}{\Lambda^{4}} &= -\frac{\left(\lambda _{M \phi  \phi  M}\right){}^3 \left(\lambda _{M \phi  \phi  M}+3 \lambda _{T \text{MM}}\right)+24 \left|\lambda
   _{\epsilon }\right|{}^2 \lambda _{M \phi  \phi  M} \left(\lambda _{M \phi  \phi  M}+6 \lambda _{T \text{MM}}\right)-128 \left|\lambda
   _{\epsilon }\right|{}^4}{96 \pi ^2 m_{Z_M}^4} \, .
\end{align}

\bibliographystyle{apsrev-title}
\bibliography{GRAVNEUTRINOBIB}

\end{document}